\documentclass[aps,twocolumn,floats,prl,psfig]{revtex4}
\input epsf
\usepackage{graphicx}

\newcommand{\beq}{\begin{equation}}
\newcommand{\beqa}{\begin{eqnarray}}
\newcommand{\eeq}{\end{equation}}
\newcommand{\eeqa}{\end{eqnarray}}

\newcommand{\lsim}{\lesssim}
\newcommand{\gsim}{\gtrsim}

\newcommand{\vect}[1]{\mbox{\boldmath${#1}$}}
\newcommand{\lmk}{\left(}
\newcommand{\rmk}{\right)}

\newcommand{\lla}{\left\langle}
\newcommand{\p}{\partial}
\newcommand{\rra}{\right\rangle}

\newcommand{\veC}{\vect C}

\begin{document}
\title{Probing the largest scale structure in the universe  with
polarization map of galaxy clusters}  

\author{Naoki Seto}
\author{Elena Pierpaoli}
\affiliation{Theoretical Astrophysics, MC 130-33, California Institute of Technology, Pasadena,
CA 91125
}
\begin{abstract}
We introduce a new formalism to  describe the polarization signal of
galaxy clusters on the whole sky. 
We show that a sparsely sampled,
 half--sky map of the cluster  polarization signal   at $z\sim 1$ 
would  allow to better characterize the very large scale density fluctuations.
While the horizon length is smaller in the past,  two  other competing
 effects  significantly remove the contribution of the small scale
 fluctuations from the  quadrupole polarization pattern at $z\sim 1$. For the standard  
$\Lambda$CDM universe with vanishing tensor mode, the quadrupole moment
 of the   
temperature anisotropy probed by WMAP  is expected to have a
$\sim32\%$ contribution  from fluctuations on scales below
 $6.3h^{-1}$Gpc.  This percentage would be reduced to
 $\sim 2\%$ level for the quadrupole moment of  polarization
 pattern  at $z\sim 1$. 
A cluster polarization map at $z \sim 1$ would 
 shed light on  the potentially anomalous features of the 
 largest scale structure in  the observable universe.
\end{abstract}
\maketitle

\underline{\em 1)  Introduction}
One of the most intriguing features of the  temperature anisotropy
 measured by WMAP \cite{Bennett:2003bz}
 is the extremely low value of the estimated quadrupole  moment. 
According to the WMAP team, the probability of
observing such a low value, given their best--fit 
cosmological model, is $\sim 0.7\%$ \cite{Spergel:2003cb}. 
Independent analysis of the WMAP data have somewhat reduced the significance of
this finding  (see {\it e.g.} \cite{Tegmark:2003ve}), but  have also
 evidenced other curious results, such as the indication of a
preferred direction for  the quadrupole and octopole modes
\cite{deOliveira-Costa:2003pu}. 
These discoveries have raised speculations 
on possible  non--standard
topology of the Universe \cite{deOliveira-Costa:2003pu,
Cornish:2003db} or 
inflationary physics (see {\it e.g.} \cite{Contaldi:2003zv}), and
at the same time,   evidenced  the importance 
of   independent observational methods to probe
 the  largest scale structure in the Universe.

The
low-$l$ moments   
of the present temperature anisotropies 
are usually assumed to probe the largest scales, but in fact 
they carry information about  perturbations on a fairly broad
 range of scales. 
It would be highly preferable to 
find an observable which   probes the largest
scales with less contamination from intermediate ones.
In this {\it Letter} we point out that the  polarization signal
from high--redshift clusters could be such observable.

Light scattered off a  free electron  
is polarized, if the electron sees a 
quadrupole anisotropy of its incident light.
This effect gives rise to interesting observable phenomena,
like, for example,
 the large--scale polarization signal in a reionized Universe
 \cite{Zalda97,Skordis}.
The same effect is responsible for the major polarization signal 
in Sunyaeve--Zeldovich (SZ) galaxy clusters.
As first pointed out by \cite{Kamionkowski:1997na} ,
the polarization pattern of clusters at different redshifts
may  
shed light on the quadrupole amplitude at earlier times and at different
positions
(see also
\cite{Baumann:2003xb} in 
relation to the low observed quadrupole moment). 
The  cluster polarization  has also  been  
investigated as a potential tool  to determine the
nature of dark energy 
that is sensitive to  the Integrated Sachs-Wolfe  effect
\cite{Baumann:2003xb,Cooray:2003hd}.

In this  {\it Letter} we develop a  transparent formalism for the calculation of
the 
relevant  
observable quantities of the cluster's polarization signal, and
show that the polarization pattern  at  $z\sim 1$ could be a promising
probe of the largest scale structure in the universe, compared with the
quadrupole temperature 
anisotropies observed today.
This finding contrasts with the naive expectation  
that the polarization signal of  distant  clusters  are generated
by fluctuations whose  spatial scales  are
well within the horizon at 
the present \cite{Port}. 
We also estimate how well the 
dark energy equation of state can be measured by the sole cluster polarization
signal.

\underline{\em 2)  Formulation}
In this section we  review the formalism to analyze  the polarization signals
of galaxy clusters.
  Throughout this  {\it Letter}, we only discuss linear  scalar
perturbations in  a flat background universe. 
We first introduce a fixed spherical coordinate system $\veC_0$  centered on
an 
observer at $z=0$.  In the coordinate system $\veC_0$  we denote the angular
variables with  $\Omega=(\theta, \phi)$ and use 
redshift $z$  as a radial coordinate for  our past
light cone. 
Let us  consider the linear polarization 
generated by the local temperature anisotropy  seen by a cluster labeled
by  $i$ 
 at a redshift $z_i$  and direction $\Omega_i$. 
Such polarization can be expressed as a linear combination of the
quadrupole  anisotropy $a_{2m}(\Omega_i,z_i)$ seen by the  cluster as \cite{Kamionkowski:1997na, Seto, Port}:
\beq
X_i=\frac{Q_i+iU_i}{F(z_i)\tau_{ci}}=\sum_{m=-2}^2 a_{2m}(\Omega_i,z_i){}_2
Y_{2m} (\Omega_i), \label{obs}
\eeq
where $F(z)=-\sqrt{6}/10  T_{CMB}(z)$ is a normalization factor
($T_{CMB}(z)$: CMB temperature 
at redshift $z$), $\tau_{ci}$ is the effective
optical depth 
of the 
cluster, and ${}_2
Y_{2m} (\Omega)$ is the spin-weighted spherical harmonics
\cite{Hu:1997hp,Zaldarriaga:1996xe} defined in the  $\veC_0$ system.  
The coefficient $a_{2m}(\Omega_i,z_i)$  in eq.(\ref{obs}) is defined in
a spherical coordinate system $\veC_i$ whose orientation is obtained by a 
 parallel transport of   $\veC_0$  to the  cluster's
position  $(\Omega_i,z_i)$.
   We shall
assume that we can make a three dimensional polarization map  
$X(\Omega,z)$ on our past light cone by observing many clusters at
different redshifts and directions.  In eq.(\ref{obs}) we only include the
primary temperature quadrupole anisotropy as the source for  
the cluster polarization, since the contribution of 
 the secondary polarized incident light 
is expected to be much weaker. 
We also neglect  the 
effects of the peculiar velocity field \cite{Kamionkowski:1997na,
Cooray:2003hd}, assuming that the typical comoving distance between the
surveyed clusters is larger than the correlation length of the peculiar
velocity field $\lsim 50 h^{-1}$Mpc \cite{gorski88}. To make the map 
$X(\Omega, z)$ we need to estimate the optical depth $\tau_{ci}$ of each
cluster. This would be performed by using  other observations like the SZ spectral distortion or the X-rays.

One way to characterize the statistical properties of the polarization
map $X(\Omega, z)$ is through its correlation function $\lla
 X(\Omega_i,z_i)  X(\Omega_j,z_j)   \rra$. This function is written in
terms of 
the correlation $\lla  a_{2m}(\Omega_i,z_i)
a_{2m'}(\Omega_j,z_j) \rra $ for the information given at two
positions $i$ and $j$  \cite{Port}.  But this representation has two disadvantages.
First, it is given by different frames of reference. Second, it is not a
diagonal matrix with respect to $m$ and $m'$. As a result its expression 
is very complicated and  hard  
to relate to basic theoretical inputs (e.g. the primordial power
spectrum).

Here, we extensively use the properties of the spin-weighted spherical
harmonics for analyzing tensor quantities. 
This kind of approach is widely used in both CMB and weak lensing
studies \cite{Hu:1997hp,Zaldarriaga:1996xe,lens}, and especially useful
for dealing with statistically isotropic fluctuations that are analyzed in this {\it
Letter}.
Our goal here is to write  the map $X(\Omega, z)$ in the orthonormal form;
\beq
X(\Omega,z)=\sum_{l=2}^\infty
\sum_{m=-l}^l{}_2Y_{lm}(\Omega)b_{lm}(z),\label{map} 
\eeq
and to present the basic formulas that relate  the coefficients
$b_{lm}(z)$ to the  spectrum $P(k)$ of the primordial density
fluctuations. 
In order to cast eq.(\ref{obs}) as in  (\ref{map}), we
  first  
separate the radial information $z$ and the directional information
$\Omega$ for the coefficient $a_{2m}(\Omega, z)$. Then
we combine the
angular    
information with that of the spin-weighted harmonics $_2Y_{2m}(\Omega)$
in eq.(\ref{obs}) 
to get the expansion with the  orthonormal angular basis ${}_2Y_{lm}(\Omega)$
as in eq.(\ref{map}).
The latter process is similar to the unification 
of the spin 
and orbital angular momentum in Quantum Mechanics ({\it e.g.}
\cite{Sakurai}), whose application to CMB is discussed in   Hu \& White
\cite{Hu:1997hp} (see also \cite{lens}).
Following  the analysis for the CMB polarization in a reionized universe 
 \cite{Hu:1997hp,Zaldarriaga:1996xe}, 
we find
\beqa
\lla  b_{lm}(z)b^*_{l'm'}(z') \rra&=&\delta_{ll'}\delta_{mm'}(4\pi)^2
\label{amp} \\
 & &\times  \int
\frac{dk}k   (P(k)k^{-1})
h_l(k,z) h_l(k,z'),\nonumber
\eeqa
where  $P(k)=Ak^n$ ($A$: a normalization factor) is the primordial
 power spectrum with  $n=1$
corresponding to  the scale invariant spectrum.   Due to the assumption
of the statistical isotropy, the correlation (3) has diagonal form.
This expression is given for E-mode (electronic parity) polarization
generated  by
scalar  perturbations that would not produce B-mode (magnetic parity)
polarization. Tensor (gravitational wave) or vector perturbations can
generate both E, and B-mode polarizations, and we can determine or 
constrain their amplitudes  by measureing B-mode
polarization. 

 In eq.(\ref{amp}) 
the function $h_l(k,z)$ is defined as
\beq
h_l(k,z)=\Delta_2(z,k) f_l[k(\tau(0)-\tau(z))],\label{hl}
\eeq
where $\tau$ is the conformal time, $\Delta_2(z,k)$ is the transfer function for the quadrupole
temperature 
anisotropies at a given $z$, and $f_l[k(\tau(0)-\tau(z))]$ is a
geometrical factor which relates the scale of the fluctuation with the
distance of the cluster  
from the observer. We will return to  this factor later on.

The transfer function $\Delta_2(z,k)$  is related to   the linear
growth rate 
$D$ and to the 
scale factor $a$ as: 
\beqa
\Delta_2(z,k)&=&\frac3{10}j_2[k(\tau(z)-\tau(z_{rec}))]\nonumber \\
& &+\frac{9}{5}\int_{\tau_{rec}}^{\tau(z)}
d\tau j_2[k(\tau(z)-\tau)]\frac{\p}{\p
\tau}\lmk\frac{D(\tau)}{a(\tau)}\rmk, \label{trans}
\eeqa
where  $j_2(x)$ is the  spherical Bessel function and 
$\tau_{rec}$ is the conformal time at recombination  $z_{rec}\sim
1100$.  
Note that the
function $\Delta_2(z,k)$ represents the 
evolution and the projection effects
of  each Fourier mode $k$,   but it  does not depend on the power
spectrum. 
The first term on the r.h.s. of eq.~(\ref{trans}) is  the Sachs-Wolfe
(SW) effect, and 
 conveys information including the  
largest scale structure (comparable to  the horizon size) at redshift $z$.
 The 
second term is the Integrated  Sachs-Wolfe (ISW) effect, and is
sensitive to the
recent expansion  history of the universe. The ISW  term  typically probes
smaller scales  than the  horizon size.

The function  $f_l(x)$  in eq.(\ref{hl})
represents projection effects for scales of the order of the cluster's distance from the observer
and it  is expressed  in terms of spherical
Bessel functions as 
$
f_l(x)\equiv\sqrt{\frac{(l+2)!}{6(l-2)!}}\frac{j_l(x)}{30 x^2}. \
$

Note that eq.(\ref{obs}) only contains the expression of
 the local quadrupole $(l=2)$
mode at the cluster's location, while the expression in 
eq.(\ref{map}) also contains  higher modes $(l\ge
3)$. This is due to the 
power transfer   from $l=2$ to $l\ge 3$ which is caused by the
spin-orbit angular momentum coupling.
 The function $f_l(x)$ regulates such transfer, 
preserving the total power ($\sum_{l=2}^\infty f_l(x)^2(2l+1)=1$).

We can now build a natural estimator of 
the total angular power spectrum for each $l$-mode  from the observed map as follows:
$
H_l(z)\equiv \sum_{m=-l}^l |b_{lm}(z)|^2.
$
We assume that the primordial potential fluctuations are random Gaussian
distributed. Then,  it is straightforward to
calculate 
the covariance of the spectrum 
\beq
Cov(H_l(z),H_{l'}(z'))=2\delta_{ll'} /(2l+1)\lmk\sum_m \lla  b_{lm}(z)b^*_{lm}(z') \rra   \rmk^2 \label{cov},
\eeq
where $\lmk\lla  b_{lm}(z)b^*_{lm}(z') \rra   \rmk$ is given by
eq.(\ref{amp}).  At $z=z'$ this expression trivially allows to evaluate
the  
cosmic variance for $H_l(z)$, leading to the familiar expression
 $Cov(H_l(z),H_{l}(z)={2/(2l+1)} H_l(z)^2$.

\underline{\em 3)  Results}
In this   section we present the general features of 
the power spectrum $H_l(z)$.
We first discuss how the observed quadrupole spectrum $H_2(z)$ is related
to the 
matter fluctuations at different wave number $k$. 
As shown in eq.(\ref{amp}) the term $h_2(k,z)^2$ represents the weight for  the
function $P(k)/k$ which  is constant for the scale 
invariant model. Therefore, we shall use $h_2(k,z)$ as 
a  measure of the spatial scale probed by  the spectrum  $H_2(z)$. 

In figure 1 we plot the function $h_2(k,z)$ at three different epochs
$z=0$, 0.5 and 1 for our fiducial cosmological parameters
$\Omega_m=0.3$ and 
$\Omega_\lambda=0.7$.   
Note that the curve for  $z=0$ can also be a regarded as  the 
weight for  the quadrupole
moment of the temperature anisotropies observed today. This curve shows that 
the temperature quadrupole receives a contribution from fluctuations
on  a broad range of scales $\sim 10^{-4}h$Mpc$^{-1}$
to 
$\sim 10^{-2}h$Mpc$^{-1}$.  The wiggles in the curve reflect the oscillatory behavior 
 of the spherical Bessel function in the first term on the r.h.s. 
of  eq.(\ref{trans}) (related to 
the SW effect), while  the negative mean value
around $10^{-3}h$Mpc$^{-1} \lsim k
\lsim 10^{-2}h$Mpc$^{-1}$ is due to the second term (ISW effect).

Let us discuss the redshift evolution  of the function $h_2(k,z)$. 
 The wave number of the
first peak of the curves $h_2(k,z)$ is determined by the horizon scale
at each redshift. 
As expected, such wave number increases as we move to higher redshift.
This scale, however, does not change much at  $z \lsim 1$:
for our fiducial cosmological model the comoving length of the  
horizon is reduced only by $14\%$  ($ 24\%$)
 at $z=0.5$ ($z=1$)  compared with the horizon size at $z=0$.

Let us now discuss  two other important effects that tend to increase the
weight of large 
spatial scales  probed by  
$H_2 (z)$ at  redshift $z \sim 1$.
 The first effect is due to the presence of  dark energy. 
 In general, our Universe
becomes  
very close to the Einstein de-Sitter one  at a relatively low
redshift $z\sim 1$. 
As a consequence, the weight associated with scales
 smaller than the horizon
(which are the ones  typically affected by the ISW effect)
is significantly reduced at $z\sim 1$.

The other  important effect that suppresses  $h_2(k,z=1)$ 
 at 
small scale   is the transfer of power from $l=2$
mode to higher modes. 
This effect is characterized by the function
$f_2(x)$ that has a profile $f_2(x)\sim 1$ at $x\lsim 1$ and $f_2(x)\sim
-15\sin(x)/x^3$ at 
$x\gsim 2$.
  The typical scale below which this suppression occurs
 is proportional to the distance to the cluster.  However, as we are
dealing with our past light cone,  this
distance coincides with  
 the  difference in the horizon sizes at the present and at 
 the cluster's redshift.
As noted above, this  length is 24 $\%$
of the present horizon for clusters at  $z= 1$,  so that  fluctuations 
 on scales smaller than  this one cannot make a important contribution to the 
observed quadrupole moment $H_2(z=1)$.
The present result is 
 straightforward  in our new formalism, while it has been 
overlooked in previous works \cite{Port} because of the complications
introduced by  
the use of multiple coordinate systems (see also \cite{Skordis}).  
From figure 1, it is apparent that the polarization map around $z=1$ would
be a powerful tool to study the largest scale fluctuations avoiding the 
small scale contamination.

\begin{figure}
  \begin{center}
\epsfxsize=8.cm
\begin{minipage}{\epsfxsize} \epsffile{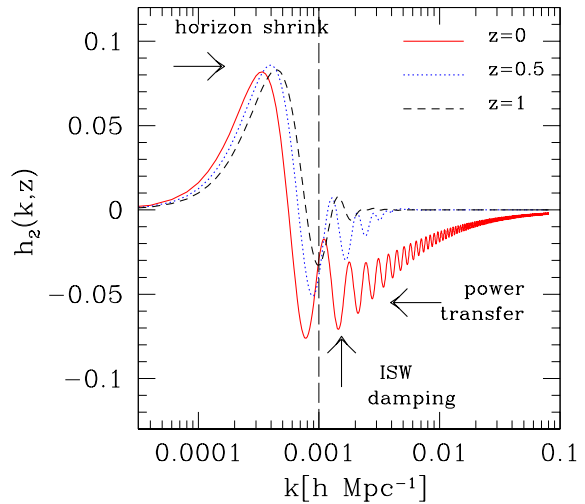} \end{minipage}
 \end{center}
  \caption{ Redshift evolution of the function $h_2(k,z)$ for the fiducial
  cosmological model with  $\Omega_m=0.3$ and 
 $\Omega_\lambda=0.7 $. Overall scale
 is irrelevant here. There are
 three effects that change the contribution of $k$-mode to the
 quadrupole pattern $H_2(z)$. Arrows show how these effects change the
 shape of the curve $h_2(k,z)$ with increasing $z$.}
\label{f1}
\end{figure}

We will now attempt a 
   quantitative analyses aimed  to determine  which wave number $k$ provides 
the dominant contribution to the function $h_l(k,z)$.
We define  a function $R(z)$
 that is the mean of $ \log_{10}[k]$
weighted by the weight $h_2(k,z)^2$  as follows
$
R(z)=\frac{\int
\frac{dk}k   \log_{10}[k]~
h_2(k,z)^2}{\int
\frac{dk}k   
h_2(k,z)^2}.
$
If we define  $\log_{10}[k_*] \equiv R(z)$, the wavelength $k_*$ can be
regarded as a typical scale probed by the moment $H_2(z)$. 
We  obtained $R(0)=-3.16$, $R(0.6)=-3.42$, $R(1.0)=-3.41$ and
$R(2.0)=-3.38$ (with $k$ is in units of $h{\rm Mpc^{-1}}$).  Roughly
speaking, we can  
reduce the effective wave number by a factor of $\sim 2$ by  using a
map at 
$z\sim 1$, compared with the quadrupole temperature anisotropies observed
today.    
For the sake of comparison we also calculated the same kind of quantity
for the temperature anisotropies at $z=0$,
and found 
$R=-3.08$ ($R=-3.00$) for $l=3$ ($l=4$).

In addition to $R(z)$, we also studied the contributions of fluctuations below
and above 
$k=10^{-3} h{\rm  Mpc^{-1}}$  to the spectrum $H_2(z)$. In figure 1 the
wave number $k=10^{-3}h{\rm 
Mpc^{-1}}$ is given by the vertical long-dashed line, and corresponds to
$\sim 6.3h^{-1}$Gpc in real scale.  At $z=0$ about
$32\%$ of the 
power is  coming from $k>10^{-3}h{\rm Mpc^{-1}}$, but its contribution
decreases to 
$\sim 2\%$ at $z=1$.
We conclude that the quadrupole of the cluster polarization signal
 at $z \sim 1$ would allow to probe the large scale power of the
Universe  in a cleaner way.

So far, we  mainly discussed the quadrupole mode. As we commented
earlier, we expect
 higher modes $(l\ge 3)$ to be  generated by the spin-orbit coupling. 
In figure 2 we show the spectrum $H_l(z)$ as a function of  redshift. 
At
$z= 0$ the polarization map $X$ simply reflects our local quadrupole
mode so that  $H_{l\ge 3}=0$. The total power $\sum_{l=2}^\infty H_l(z)$ (dashed curve)  is 
 the averaged local temperature quadrupole moment $\sum_m\lla
a_{2m}a^*_{2m} \rra  $ at each 
redshift (see eq.(\ref{obs})).  At $z\gsim 
1$ it becomes a constant value due to the scale invariance of the matter
power spectrum. The quadrupole mode $H_2(z)$ decreases
continuously with increasing redshift due to 
the power transfer to higher-$l$ modes.
However, the $l=2$ mode remains the largest signal at  redshifts
$z\lsim 
2$ where the cluster polarization map would be observationally
available. 

\begin{figure}
  \begin{center}
\epsfxsize=8.cm
\begin{minipage}{\epsfxsize} \epsffile{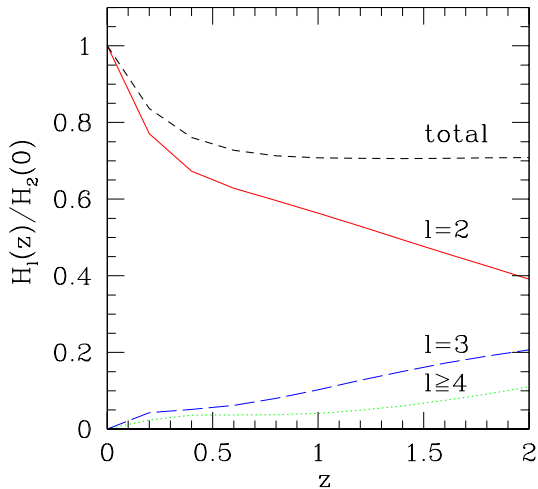} \end{minipage}
 \end{center}
  \caption{ Redshift evolution of the spectrum $H_l(z)$ for parameters  $\Omega_m=0.3$,
 $\Omega_\lambda=0.7 $ and  $n=1$.}
\label{f1}
\end{figure}

We shall now consider  if the redshit dependence of the polarization 
signal can be used to constrain cosmology. In particular we 
investigate possible constraints on the equation of state of dark energy.
We evaluate the parameter estimation errors using
 the  Fisher matrix approach applied to $H_2(z)$. 
We combine information on $H_2(z)$ from different redshift shells
between $0\le
z \le 2 $ binned 
in
$\delta z=0.2$ intervals, and assume  cosmic
variance with appropriate bin correlations 
(see  eq.(\ref{cov}), and also \cite{Port}),
 as the sole source of the error.

We constrain a single fitting parameter
$w=P/\rho=$constant around the fiducial model ($\Omega_m=0.3$,
$\Omega_\lambda =0.7$ and $n=1$),
and find $\Delta w\simeq 0.6$ ($1\sigma$). 
Such a large error is due to the large cosmic variance in each redshift
bin 
and to the strong correlations between bins.
Indeed the SW effect is not important for  dark energy studies,
but it  contributes to the above correlation. Thus, in the attempt of 
removing the SW contribution,  
we also applied the Fisher matrix approach to  the estimator
$\sum_m\lla| b_{2m}(z)-b_{2m}(0)|^2\rra$,  
and found
$\Delta w\simeq 0.4$ ($1\sigma$).
In order to improve this result we could  also include, in principle,
 information of the higher order modes $l\ge 3$.
In an actual observational analysis, we need a lot of efforts to  detect
these weak signals around  at $z\lsim 1$ where the effect of
the dark energy is important (see figure 2). 
We conclude that cluster polarization alone would not be 
a powerful observational method to constrain dark energy.

Finally let us make  an order of magnitude estimate for observational
requirement  needed to measure the moment $H_2(z)$ at $z\sim 1$ (see also
\cite{Cooray:2003hd}). Suppose we 
 measure the polarization signal for  $N$ clusters distributed on the whole sky at
$z\sim 1$ with an observational error $\sigma$ on the polarization intensity
 for each cluster.
 The total error for  $H_l$ is  $\Delta
H_l=\sqrt{2(2l+1)}[H_l/(2l+1)+4\pi \sigma^2/N]$ with the first term
representing the cosmic variance and the second term being the measurement
error. For $l=2$ the two errors are  equal  if
$\sigma\sim 
0.3(\tau_c /10^{-2})(N/10^3)^{1/2}\mu$K.
A part from the sensitivity considerations, there is also 
the issue of separating the cluster polarization signal
 from the other competing ones, like, for example, the  CMB lensing.
This is best achieved if the cluster is spatially resolved, which typically 
requires observations with resolution 
of $\sim 1$ arcmin. 
Planck, for example, is an all sky survey mission with sensitivity $\sim
5\mu$K per pixel (for combined polarization channels and 14 months integration)
 and  $\sim 5$ arcmin angular resolution.
Because of the low resolution, it may not be the 
most suited experiment  to detect $H_2$ around $z\sim1$.


As the polarization pattern 
is dominated by the low-$l$ modes, a fine sky sampling is not necessary to map it.
 Furthermore, in order  to probe only $l=2$ mode at $z\lsim 1$ where 
$l\ge 3$ modes are weak, we 
just need to observe
half of the sky  ($2\pi$[sr]) due to the parity symmetry of scalar
perturbations (though, in this case,  the power transfer in figure 1
does not work).   
A ground--based telescope performing targeted cluster observations
in small areas  sparsely distributed on half sky may be an adequate 
tool to achieve this task. 

The authors would like to thank A. Cooray and M. Sasaki for discussions.
N.S. is supported by NASA grant NNG04GK98G and  the Japan Society
for the 
Promotion of Science.
E.P. is an ADVANCE fellow (NSF grant AST-0340648) and  also supported by NASA grant NAG5-11489.


\begin{thebibliography}{99}
%

\bibitem{Bennett:2003bz}
  C.~L.~Bennett {\it et al.},
  Astrophys.\ J.\ Suppl.\  {\bf 148}, 1 (2003).

\bibitem{Spergel:2003cb}
  D.~N.~Spergel {\it et al.},
  Astrophys.\ J.\ Suppl.\  {\bf 148}, 175 (2003).


\bibitem{Tegmark:2003ve}
  M.~Tegmark,  {\it et al.},
  Phys.\ Rev.\ D {\bf 68}, 123523 (2003);
  G.~Efstathiou,
  Mon.\ Not.\ Roy.\ Astron.\ Soc.\  {\bf 346}, L26 (2003);
 I.~J.~O'Dwyer {\it et al.},
  Astrophys.\ J.\  {\bf 617}, L99 (2004).



\bibitem{deOliveira-Costa:2003pu}
A.~de Oliveira-Costa, M.~Tegmark, M.~Zaldarriaga and A.~Hamilton,  
  Phys.\ Rev.\ D {\bf 69}, 063516 (2004);


\bibitem{Cornish:2003db}
 N.~J.~Cornish  {\it et al.},
  Phys.\ Rev.\ Lett.\  {\bf 92}, 201302 (2004).



\bibitem{Contaldi:2003zv}
  C.~R.~Contaldi   {\it et al.},
  JCAP {\bf 0307}, 002 (2003).





\bibitem{Zalda97}
  M.~Zaldarriaga, D.~N.~Spergel and U.~Seljak,
  Astrophys.\ J.\  {\bf 488}, 1 (1997);
  O.~Dore, G.~P.~Holder and A.~Loeb,
  Astrophys.\ J.\  {\bf 612}, 81 (2004).


\bibitem{Skordis}
  C.~Skordis and J.~Silk,
  arXiv:astro-ph/0402474.



\bibitem{Kamionkowski:1997na}
M.~Kamionkowski and A.~Loeb,
Phys.\ Rev.\ D {\bf 56}, 4511 (1997).


\bibitem{Baumann:2003xb}
  D.~Baumann and A.~Cooray,
  New Astron.\ Rev.\  {\bf 47}, 839 (2003).



\bibitem{Cooray:2003hd}
  A.~Cooray,  {\it et al.},
  Phys.\ Rev.\ D {\bf 69}, 027301 (2004).



\bibitem{Port}
J.~Portsmouth,
Phys.\ Rev.\ D {\bf 70}, 063504 (2004).


\bibitem{Seto}
N.~Seto and M.~Sasaki,
Phys.\ Rev.\ D {\bf 62}, 123004 (2000);
 A.~Amblard and M.~J.~White,
  arXiv:astro-ph/0409063.




\bibitem{Hu:1997hp}
W.~Hu and M.~J.~White,
Phys.\ Rev.\ D {\bf 56}, 596 (1997).


\bibitem{Zaldarriaga:1996xe}
M.~Zaldarriaga and U.~Seljak,
Phys.\ Rev.\ D {\bf 55}, 1830 (1997);
M.~Kamionkowski, A.~Kosowsky and A.~Stebbins,
  Phys.\ Rev.\ D {\bf 55}, 7368 (1997).

\bibitem{gorski88}
 K.~Gorski,
  Astrophys.\ J.\  {\bf 332}, L7 (1988).

\bibitem{lens}
  P.~G.~Castro, A.~F.~Heavens and T.~D.~Kitching,
  arXiv:astro-ph/0503479.





\bibitem{Sakurai}
J. J. Sakurai, {\it Modern Quantum Mechanics} (Addison-Wesley, New York, 1985).

\end{thebibliography}
\end{document}